\documentstyle[fleqn,gc,epsf]{article}

\def\Jl#1#2{{\it #1\/} {\bf #2},\ }

\def\GC#1 {\Jl{Grav. \& Cosmol.}{#1}}
\def\JMP#1 {\Jl{J. Math. Phys.}{#1}}
\def\PRD#1 {\Jl{Phys. Rev.}{D\ #1}}
\def\PRL#1 {\Jl{Phys. Rev. Lett.}{#1}}
\def\PLB#1 {\Jl{Phys. Lett.}{#1B}}

\begin{document}

\hyphenation{cos-mo-logy} \hyphenation{asymp-to-tic}
\hyphenation{sti-mu-lating} \twocolumn[

\Title{WORMHOLES IN THE BRANEWORLD} \Author{Makarenko A.N.} {Lab.
for Fundamental Study, Tomsk State Pedagogical University, Russia}

\Abstract{ We discuss brane wormhole solution when classical brane
action
 contains 4d
 curvature. The  equations of motion for the
 cases with $R=0$ and $R\ne 0$ are obtained. Their numerical solutions
corresponding to wormhole are found for specific boundary
conditions.  } ]

The wormhole is very important object in modern cosmology. It is
suggested that wormholes are bridges linking two distinct
spacetimes, or handles between remote parts of a single universe,
see for example \cite{l1},\cite{l3},\cite{l4}. Wormhole is the
solution of the Einstein equation if the stress-energy tensor of
matter violates the null energy conditions (NEC) \cite{l4}. There
are different ways of evading these violations. Most of these
attempts focus on alternative gravity theories or existing of
exotic matter \cite{l5}, \cite{l6}, \cite{l7}. We consider the
wormholes in frames of brane worlds \cite{l8}. The brane world
scenario assumes that our Universe is four-dimensional spacetime,
embedded in the 5D bulk spacetimes. According to this concept the
4d Einstein equations will be modified if we use the Gauss and
Codacci equations \cite{l9}. The purpose of this note is to
construct brane wormholes in 5d space.

Let $g_{\mu\nu}$ be the metric of the bulk space and $n_{\mu}$ is
the unit vector normal to the 3-brane. Then metric induced on the
brane has the form:
\begin{equation}
q_{\mu\nu}=g_{\mu\nu}-n_\mu n_\nu.
\end{equation}

We start from action:
\begin{eqnarray}
S=\int d^5 x\sqrt{-g}\left[\frac{1}{k_5^2}
R^{(5)}-2\Lambda\right]+S_{brane}(q).
\end{eqnarray}
Here suffix (5) denotes the 5d quantities. The $\Lambda$ is 5d
cosmological constant and $S_{brane}(q)$ is the action on the
brane. The bulk Einstein equation is given by
\begin{eqnarray}
\frac{1}{k_5^2}\left(R^{(5)}_{\mu\nu}-\frac{1}{2}g_{\mu\nu}R^{(5)}\right)=T_{\mu\nu}.
\end{eqnarray}
We assume the 5d metric to have the form
\begin{equation}
ds^2=d\xi^2+q_{\mu\nu}dx^\mu dx^\nu.
\end{equation}
Then the 5d stress-energy tensor takes the following form
\begin{equation}
T_{\mu\nu}=-\Lambda g_{\mu\nu}+(-\lambda
q_{\mu\nu}+\tau_{\mu\nu})\delta(\xi).
\end{equation}
Here $\lambda$ is 4d cosmological constant and $\tau_{\mu\nu}$
represents the contribution due to brane matter.

In this case the brane Einstein equation can be represented in the
form

\begin{eqnarray}
&&\frac{1}{k_5^2}G^{(4)}_{\mu\nu}=
-\frac{1}{2}\left(\Lambda+\frac{k_5^2\lambda^2}{6}\right)q_{\mu\nu}+\nonumber\\
&&\frac{k_5^2\lambda}{6}\tau_{\mu\nu}+k_5^2
\pi_{\mu\nu}-\frac{1}{k_5^2}E_{\mu\nu}.
\end{eqnarray}
Here $E_{\mu\nu}$ is the part of the 5d Weyl tensor defined by
$$E_{\mu\nu}=C^{(5)}_{\alpha\beta\gamma\delta}n^\alpha n^\beta
q_\mu^\gamma q_\nu^\delta,$$ and $\pi_{\mu\nu}$ is given by
\begin{eqnarray}
\pi_{\mu\nu}=-\frac{1}{4}\tau_{\mu\alpha}\tau_\nu^\alpha+\frac{1}{12}\tau\tau_{\mu\nu}+
\frac{1}{8}q_{\mu\nu}\tau_{\alpha\beta}\tau^{\alpha\beta}-\frac{1}{24}q_{\mu\nu}\tau^2.\nonumber
\end{eqnarray}

Let us consider the brane action to be the following form

\begin{equation}
S_{brane}=\int d^4x\sqrt{-q}\left(-\alpha
R^{(4)}(q)-2\lambda\right).
\end{equation}
Here $\alpha$ is arbitrary parameter. In this case the brane
equation is taken as following \cite{l10}:

\begin{eqnarray}
\label{eqn2}
&&\frac{1}{\kappa_5^2}\left(1
-\frac{k_5^4\lambda\alpha}{6}\right)\left(R^{(4)}_{\alpha\beta}
-\frac{R^{(4)}}{2}q_{\alpha\beta}\right)=\nonumber\\
&&-\frac{1}{2}\left(\Lambda+\frac{k_5^2\lambda^2}{6}\right)q_{\alpha\beta}
+\alpha^2\kappa_5^2\left[-\frac{1}{4}R^{(4)}_{\alpha\mu}
R^{(4)\mu}_\beta+\right.\nonumber\\
&& \left.\frac{1}{6}R^{(4)}R^{(4)}_{\alpha\beta}-
q_{\alpha\beta}\left(\frac{1}{16}
{R^{(4)}}^2-\frac{1}{8}R^{(4)}_{\alpha\beta}
R^{(4)\alpha\beta}\right)\right]-\nonumber\\
&&\frac{1}{\kappa_5^2}C^{(5)}_{\mu\nu\rho\delta}n^\mu n^\rho
q^\nu_\alpha q^\delta_\beta.
\end{eqnarray}

Let us consider the static, spherically symmetric metric on the
brane:
\begin{eqnarray}
\label{m1}
&&ds^2=-e^{2a_1(r)}dt^2+e^{2a_2(r)}dr^2+\nonumber\\
&&r^2(d\theta^2+\hbox{Sin}^2\theta\; d\varphi^2).
\end{eqnarray}

The 4d curvature for metric (\ref{m1}) has the form
\begin{eqnarray}
&&R^{(4)}=\frac{2e^{-2a_2}}{r^2}\left(-1+e^{2a_2}-r^2{\dot{a}_1}^2+\right.\nonumber\\
&&\left.
2r{\dot{a}_2}+r{\dot{a}_1}(-2+r{\dot{a}_2})-r^2\ddot{a}_1\right).
\end{eqnarray}

It is simple to find the 4d combination
\begin{eqnarray}
\label{eqn3}
&&R^{(4)}_{\alpha\beta}{R^{(4)}}^{\alpha\beta}=\nonumber\\
&&e^{-4a_2}\left(\frac{2\left[-1+e^{2a_2}-r{\dot{a}_1}+r{\dot{a}_2}\right]^2}{r^4}+
\right.\nonumber\\
&&\left[-{\dot{a}_1}^2+\frac{2{\dot{a}_2}}{r}+{\dot{a}_1}{\dot{a}_2}
-\ddot{a}_1\right]^2+\\
&&\left.\frac{\left[r{\dot{a}_1}^2+{\dot{a}_1}(2-r{\dot{a}_2})+
r\ddot{a}_1\right]^2}{r^2}\right)\nonumber
\end{eqnarray}

If we consider the situation when $R^{(4)}=0$, one  obtaine the
equation:
\begin{eqnarray}
\label{eqn5}
&&\ddot{a}_1=\nonumber\\
&&\frac{-1+e^{2a_2}-r^2{\dot{a}_1}^2+2r{\dot{a}_2}+
r{\dot{a}_1}(-2+r{\dot{a}_2})}{r^2}.
\end{eqnarray}

Now we rewrite the Eq. (\ref{eqn3}) in the form:
\begin{equation}
\label{eqn6}
R_{\alpha\beta}R^{\alpha\beta}=
\frac{e^{-4a_2}}{r^4}\left(\frac{3}{2}[b_1^2+b_2^2]+b_1
b_2\right).
\end{equation}

Here $b_1=-1+e^{2a_2}+2r {\dot{a}_2},\;\;\;b_2=-1+e^{2a_2}-2r
{\dot{a}_1}$.

Since the Weyl tensor is traceless:
\begin{equation}
\label{eqn7} R^{(4)}_{\alpha\beta}R^{(4)\alpha\beta}=
\frac{8}{\alpha^2}\left(\frac{\Lambda}{\kappa_5^2}+\frac{\lambda^2}{6}\right).
\end{equation}

If $\Lambda$ is zero and $\lambda$ is zero, we obtain the
Schwarzschild metric, but this metric is Richi flat:
\begin{eqnarray}
&&ds^2=-c_1\left(\frac{1}{c_2}-\frac{1}{r}\right)dt^2
+\frac{dr^2}{1-\frac{c_2}{r}}+\nonumber\\
&&r^2(d\theta^2+\hbox{sin}^2\theta\; d\varphi^2).\nonumber
\end{eqnarray}

It is easy to see, that if $\lambda$ is zero, then $\Lambda$ one
should be positive.

One can see, that the Ricchi tensor have the form:
\begin{eqnarray}
R^{(4)\alpha}_\beta=\frac{e^{-2a_2}}{r^2}\hbox{diag}\left(-b_1,\;-b_2,
\;\frac{b_1+b_2}{2},\;\frac{b_1+b_2}{2}\right).\nonumber
\end{eqnarray}

The Eq. (\ref{eqn3}) can be rewritten in the form:
\begin{eqnarray}
&&\frac{1}{\kappa_5^2}\left(1
-\frac{k_5^4\lambda\alpha}{6}\right)R^{(4)\mu}_\alpha q_{\mu\beta}
=
\frac{1}{2}\left(\Lambda+\frac{k_5^2\lambda^2}{6}\right)q_{\alpha\beta}-\nonumber\\
&&\frac{\alpha^2\kappa_5^2}{4} R^{(4)\mu}_\alpha q_{\mu\nu}
R^{(4)\nu}_\beta-\frac{1}{\kappa_5^2}E_{\alpha\beta}.\nonumber
\end{eqnarray}

The classical Einstein equation in 4 dimensions is taken as
following:

\begin{equation}
R^{(4)}_{\alpha\beta}-\frac{1}{2}R^{(4)} q_{\alpha\beta}=k_4^2
T_{\alpha\beta},
\end{equation}
Or
\begin{equation}
R^{(4)}_{\alpha\beta}=k_4^2 T_{\alpha\beta}.
\end{equation}

Hence the stress-energy tensor looks like:
\begin{eqnarray}
&&T_{\alpha\beta}=\kappa_5^2{\left(1
-\frac{k_5^4\lambda\alpha}{6}\right)}^{-1}\left(
\frac{1}{2}\left(\Lambda+\frac{k_5^2\lambda^2}{6}\right)q_{\alpha\beta}\right.+\nonumber\\
&&\left.-\frac{\alpha^2\kappa_5^2}{4} R^{(4)\mu}_\alpha q_{\mu\nu}
R^{(4)\nu}_\beta-\frac{1}{\kappa_5^2}E_{\alpha\beta}\right).\nonumber
\end{eqnarray}

In order to find the solution of the Eqs. (\ref{eqn5}-\ref{eqn7})
let us consider the following of $b_1$ and $b_2$
\begin{eqnarray}
b_1&=&r^2 e^{2a_2} f_1(r),\\
b_2&=&r^2 e^{2a_2} f_2(r).
\end{eqnarray}

Then we obtain the equation on the function $f_1(r)$ and $f_2(r)$
\begin{eqnarray}
\label{ee1}
&&\frac{3}{2}(f_1(r)^2+f_2(r)^2)+f_1(r) f_2(r)=\beta^2\\
\label{ee2}
&&-2c+r+r^5f_2(r)^2-2cr^3\dot{f}_2(r)-3c\dot{f}(r)+r\dot{f}(r)+\nonumber\\
&&f(r)[3+2r^2\dot{f}_2(r)
+3\dot{f}(r)]-\nonumber\\
&&r^2f_2(r)[5c+2r-5f(r)+r\dot{f}(r)]=0,\\
&&f_1(r)=\frac{1+\dot{f}(r)}{r^2}.
\end{eqnarray}
Were c is the constant of integration and
$\beta^2=\frac{8}{\alpha^2}\left(\frac{\Lambda}{\kappa_5^2}+\frac{\lambda^2}{6}\right)$.

The Eqs. (\ref{ee1}) and (\ref{ee2}) can be solved numerically.
The explicit numerical solution is given by figures 1 and 2.

\begin{figure}
\epsffile{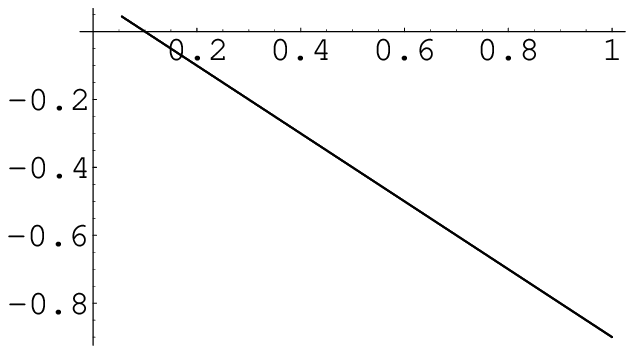} \caption{Evolution of the function $f(r)$. }
\label{figpot} \epsffile{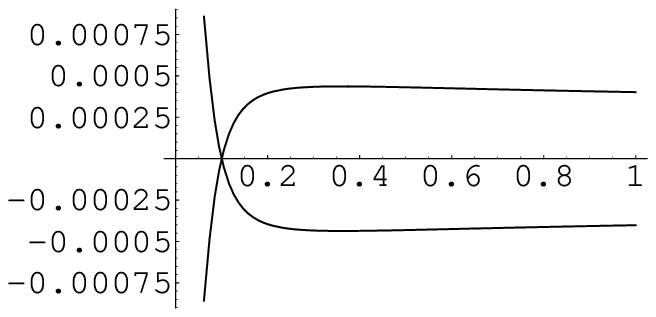} \caption{Evolution of the
function $f_2(r)$.Here $c=1,\;\;\beta=0.001$.} \label{figpot}
\end{figure}

If the condition $R=0$ is not satisfied, then one obtains the
equation:
\begin{eqnarray}
\label{jj1} &&\frac{1}{k_5^2}\left(1-\frac{k_5^2
\lambda\alpha}{6}\right)R^{(4)}-2\left(\Lambda+\frac{k_5^2\lambda^2}{6}\right)+\nonumber\\
&&\alpha^2
k_5^2\left(\frac{1}{4}R^{(4)}_{\alpha\beta}R^{(4)\alpha\beta}-\frac{1}{12}R^{(4)2}\right)=0.
\end{eqnarray}

As an example, we discuss the situation when $a_1=0$. In this case
the equation (\ref{jj1}) accepts the form:
\begin{eqnarray} &&-\alpha^2\left(e^{2a_2}-1\right)k_5^4+2\alpha
e^{2a_2}\left(e^{2a_2}-1\right)k_5^2\lambda
r^2+\nonumber\\
&&2e^{2a_2}r^2\left(6+e^{2a_2}\left[-6+k_5^4\lambda^2r^2+\right.\right.\nonumber\\
&& \left.\left. 6k_5^2\Lambda
r^2\right]\right)+2r\left(\alpha^2\left(e^{2a_2}-1\right)k_5^2-\right.\nonumber\\
&&\left. 12e^{2a_2}r^2+2\alpha e^{2a_2} k_5^2\lambda
r^2\right)\dot{a}_2-\alpha^2k_5^4 r^2\dot{a}_2^2=0.\nonumber
\end{eqnarray}
This equation can be solved numerically, as is drawn in figure 3.
\begin{figure}
\epsffile{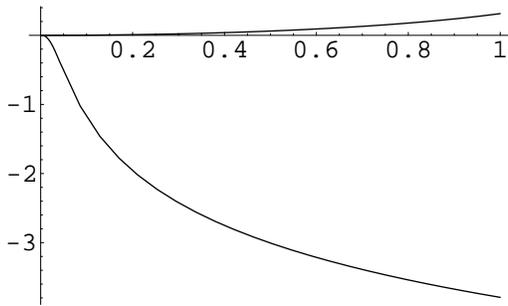} \caption{Evolution of the function $a_2(r)$.
All constants are equal to one.} \label{figpot} \label{figpot}
\end{figure}

Note if the bulk space is (Anti) deSitter  (Weyl tensor is zero),
one can obtain the following solution which describe 4d universe
with constant curvature:
\begin{equation}
R^{(4)}_{\alpha\beta}=h q_{\alpha\beta},\;\;R^{(4)}=4 h.
\end{equation}
\begin{equation}
h=\frac{6-\alpha\kappa_5^4\lambda\pm \sqrt{6}\sqrt{6-2\alpha\kappa_5^4\lambda-
\alpha_2\kappa_5^6\Lambda}}{\alpha^2\kappa_5^4}.
\end{equation}

 \Acknow {The author is grateful to S.D. Odintsov and K.E. Osetrin for
formulation the problem and numerous helpful discussions. The
research was supported RFBR No 03-01-00105 and grant for Leading
Scientific Schools, project No 1252.2003.2.}

\end{document}